\input{aipcheck}

\documentclass[
    ,final            
  ]
  {aipproc}

\layoutstyle{8x11double}

\begin{document}

\title{$\gamma$\,Doradus stars in the COROT exoplanets fields: \\ first inspection}

\classification{97.30.Dg}
\keywords      {$\gamma$\,Doradus -- frequency analysis -- fundamental parameters}

\author{P. Mathias}{
  address={UMR 6525 H. Fizeau, UNS, CNRS, OCA, Campus Valrose, F-06108 Nice Cedex 2, France}
  ,altaddress={Laboratoire d'Astrophysique de Toulouse-Tarbes, 
   Universit\'e de Toulouse, CNRS, 57 avenue d'Azereix, F-65000 Tarbes, France}
  ,email={Philippe.Mathias@oca.eu}
}
\author{E. Chapellier}{
  address={UMR 6525 H. Fizeau, UNS, CNRS, OCA, Campus Valrose, F-06108 Nice Cedex 2, France}
}
\author{M. Bouabid}{
  address={UMR 6525 H. Fizeau, UNS, CNRS, OCA, Campus Valrose, F-06108 Nice Cedex 2, France}
  ,altaddress={Institut d'Astrophysique et de G\'eophysique, Universit\'e de Li\`ege, 
   All\'ee du 6 Ao\^ut 17, B-4000 Li\`ege, Belgique}
}
\author{E. Rodriguez}{
  address={Instituto de Astrofisica de Andalucia, CSIC, Apdo. 3004, 18080 Granada, Spain}
}
\author{E. Poretti}{
  address={INAF-Osservatorio Astron. di Brera, Via E. Bianchi 46, I-23807 Merate (LC), Italy}
}
\author{M. Paparo}{
  address={Konkoly Observatory of the Hungarian Academy of Sciences, P.O. Box 67, H-1525 Budapest, Hungary}
}
\author{M. Hareter}{
  address={Institut f\"ur Astronomie, Universit\"at Wien T\"urkenschanzstrasse 17, A-1180 Vienna, Austria}
}
\author{P. De Cat}{
  address={Royal Observatory of Belgium, Ringlaan 3, B-1180 Brussels, Belgium}
}
\author{L. Eyer}{
  address={Observatoire de Gen\`eve, 51 ch. des Maillettes, CH-1290 Sauverny, Switzerland}
}

\begin{abstract}
We present here preliminary results concerning 32 stars identified as main $\gamma$\,Doradus candidates by
the COROT Variable Classifier (CVC) among the 4 first fields of the exoplanet CCDs.
\end{abstract}

\maketitle


\section{Introduction}

The COROT $\gamma$\,Doradus Thematic Team (\url{http://www.oca.eu/gdor_corot/index.html} contact: Philippe.Mathias@oca.eu) involves about 
40 persons in 30 institutes, brought together through the COROT Announcement of Opportunity for the Additional Programs.
The aims of this Additional Program are mainly the definition of the observational instability strip (IS), from which
new insights should be obtained concerning the destibilising mechanism (rotation, metallicity, evolution, tidal effects...), 
together with hybrid ($p$- and $g$-mode pulsators) detection.

\section{Candidates in the 4 first runs}

In February 2009 were released by the COROT Variable Classifier (CVC) the $\gamma$\,Doradus candidates detected in the first four runs: 
IRa1, SRc1, LRa1 and LRc1.
The respective duration of each of these runs was 54.7, 20.3, 130.9 and 142.1 days.

There were more than 1000 stars being classified by the CVC as possible $\gamma$\,Doradus candidates, among which 32 were first priority 
candidates, the others being balanced between 2nd and 3rd priority targets.
Note that due to the very poor spectro-photometric information, all the light curves obtained through the prism have been summed in order 
to deliver only white light curves. 

\begin{figure}
  \includegraphics[height=.3\textheight,angle=270]{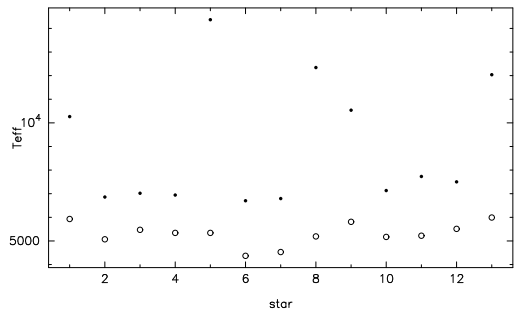}
  \caption{Comparison between the effective temperature provided in the EXODAT catalog, and that derived from Str\"omgren photometry
(C. Aerts, private communication, photometric data provided by the Binary Team).}
  \label{Fig1}
\end{figure}

Most of the stars within the COROT eyes have been observed and classified for exoplanet purpose, with results being centralized in the
EXODAT database. 
Note that this information is essential to differentiate stars having similar morphologies of their light curves; this is particularly true for
SPBs/$\gamma$\,Doradus stars that share the same typical frequency range.
However, the data in the EXODAT catalog should be treated with caution (Fig.\,\ref{Fig1}). 

It is therefore essential to complement the space data with other ones: colour photometry and 
high resolution spectroscopy. We are in particular part of the proposal 
concerning the classification of stars using ESO's FLAMES instrumenti (PI: C. Neiner, LESIA,
Meudon).

\section{Preliminary analysis of the 32 P1 candidates}

Before the frequency analysis, performed both with Period04 \cite{LB05} and SigSpec \cite{R07}, the data set is corrected from the ``flags'' 
provided by the COROT Consortium (SAA, orbital eclipses\ldots) and the hot pixels. 
These latter are particularly tricky since they produce jumps of different amplitudes associated to short-time trends related 
to the consecutive relaxation.
The resulting frequency analyses (performed until the frequency has a significance of 5 in the SigSpec frame) show many frequencies
related to the satellite orbit (in particular at 2.01\,d$^{-1}$ and 13.97\,d$^{-1}$) together with their numerous harmonics/aliases.

Once these frequencies (more than 30) are removed, there remains a group at low frequency that is usually related to the trends; these latter
are not easily taken into account by the process.
These trends represent therefore the main difficulty to deduce the 'intrinsic' stellar frequency, since trends frequencies might combine with true ones, 
even at high values.

\begin{figure}
  \includegraphics[height=.3\textheight,angle=270]{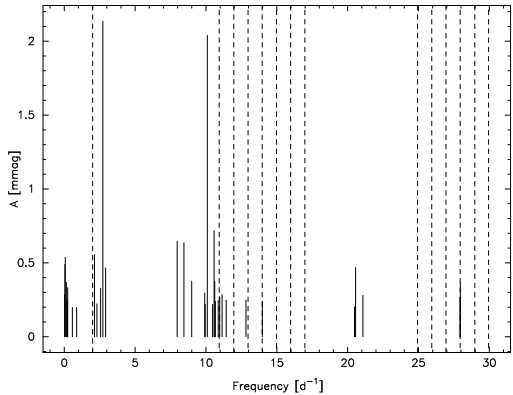}
  \caption{Frequencies of a star belonging to the IRa1 run. The red dashed-lines correspond to the satellite frequencies.
One can notice a group of very low frequencies ($< 0.05$\,d$^{-1}$) that still represent the long-term trends of the data despite
a correction using spline-functions. Finally, the remaining frequencies should be intrinsic to the star unless some might be a 
combination between a 'real' frequency and a low-one corresponding to a trend.
Note that some frequencies are in the typical $\gamma$\,Doradus domain, while others are quite high, above 20\,d$^{-1}$.}
  \label{Fig2}
\end{figure}

Figure\,\ref{Fig2} shows a typical frequency map for a hybrid $\gamma$\,Doradus star. 
The 'hybrid' status of the star is based on the presence of frequencies that are typical for $\gamma$\,Doradus stars together
with much higher frequencies, above 20\,d$^{-1}$.

\section{Results for the P1 sample}

From the light curve behaviour and the frequency analysis, we are able to classify the sample of the 32 stars in different categories:
\begin{itemize}
\item $\gamma$\,Dor candidates: 10 (+3 uncertain)
\item hybrid $g$- \& $p$-mode pulsators: 2
\item constant stars: 13. 
\item unclassified cases: 4. Among these 4, one might be a RR Lyrae star since both the basic and its 2 harmonics dominate the residuals. 
Other cases might be related to stellar activity.
\end{itemize}
Note that these results may slightly evolve through a much powerful treatment of the different trends in particular. The number of hybrid should
in particular increase, while the number of constant stars should decrease.

\begin{figure}
  \includegraphics[height=.6\textheight,angle=270]{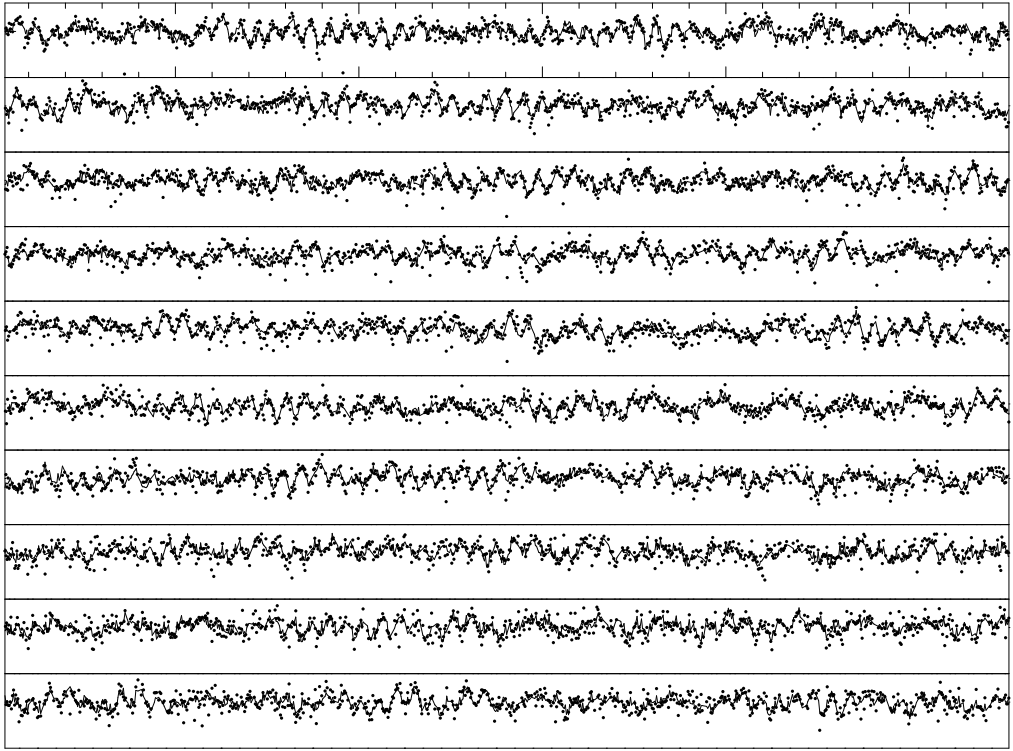}
  \caption{Sine-fit performed on the light curve corresponding to the star presented in Fig.\,2.
A total of 58 frequencies has been used (including the trends and the satellite ones), but it is clear that this fit can be improved.}
  \label{Fig3}
\end{figure}

Figure\,\ref{Fig3} shows the sine-fit of the star corresponding to Fig.\,\ref{Fig2}. 

\section{Foreseen work}

In order to manage such an enormous amount of data, it is essential to develop automated procedures to take care of the flags, 
the trends, the outliers, and, very tricky, the jumps. Up to now, only part of these tasks are done in an automated way.

However, a fundamental problem is to be sure that the star is indeed a $\gamma$\,Doradus star: the frequency spectrum itself is not 
enough to conclude to a $\gamma$\,Doradus membership (the confusion with SPB stars is obvious): complementary observations are required. 

In any case, ground based data are requested. Indeed, in addition to the location of the star in the HR Diagram, other fundamental
parameters have to be obtained in order to perform modelling and consecutive asterosismic studies.

A difficult problem, especially for $\gamma$\,Doradus stars, is the knowledge of the stellar rotation.
Indeed, what we measure is the frequency spectrum in the solar frame, that might be completely different from the real one i.e., in the
stellar frame; the fequency shift depends on the rotation period, the azimuthal number and the whole stellar structure.
Even considering $g$-modes, and a quite low $m$-value (2), a rotation frequency of 10\,d$^{-1}$ may shift the considered frequency by 10\,d$^{-1}$!
Therefore, the key-objects that are the hybrid stars have to be validated in the correct frame. Thus, the case given in this paper may also
be interpreted through a pure $g$-mode pulsator.

Therefore, whereas space missions will bring fundamental insights on the precision and the quality of the temporal signal, it is
essential to accompany them with complementary observations, especially in order to derive the fundamental parameters of the stars.

\end{document}